\documentclass[12pt]{article}
\usepackage{epsf} 
\usepackage{amsmath}
\usepackage{amssymb}
\usepackage{epsfig}
\usepackage{latexsym}
\usepackage{amsfonts}
\usepackage{graphicx}%
\usepackage{varioref}
\usepackage{ifthen}
\begin{document}
\parindent 0mm 
\setlength{\parskip}{\baselineskip} 
\thispagestyle{empty}
\pagenumbering{arabic} 
\setcounter{page}{0}
\mbox{ }
\rightline{UCT-TP-269/07}
\newline
\rightline{MZ-TH/07-15}
\newline
\rightline{November 2007}
\newline
\rightline{Revised February 2008}
\newline
%\vspace{0.1cm}
\begin{center}
{\large {\bf 	Strange quark   condensate from QCD sum rules to five loops}}
{\LARGE \footnote{{\LARGE {\footnotesize Supported in part by  NRF (South Africa) and DFG (Germany).}}}}
\end{center}
\vspace{.05cm}
\begin{center}
{\bf Cesareo A. Dominguez}$^{(a)}$, {\bf Nasrallah F. Nasrallah},$^{(b)}$, {\bf Karl Schilcher},$^{(c)}$
\end{center}
\begin{center}
$^{(a)}$Centre for Theoretical Physics and Astrophysics\\[0pt]University of
Cape Town, Rondebosch 7700, South Africa\\
$^{(b)}$ Faculty of Science, Lebanese University, Tripoli, Lebanon\\
$^{(c)}$ Institut f\"{u}r Physik, Johannes Gutenberg-Universit\"{a}t\\
Staudingerweg 7, D-55099 Mainz, Germany
\end{center}
%\vspace{0.2cm}
\begin{center}
\textbf{Abstract}
\end{center}
\noindent
It is argued that it is valid to use QCD sum rules to determine the scalar and pseudoscalar two-point functions at zero momentum, which in turn determine the ratio of the strange to non-strange quark condensates 
$R_{su} = \frac{<\bar{s} s>}{<\bar{q} q>}$ with ($q=u,d$). This is done in the framework of a new set of QCD Finite Energy Sum Rules (FESR) that involve as integration kernel a second degree polynomial, tuned to reduce considerably the systematic uncertainties in the hadronic spectral functions. As a result, the parameters limiting the precision of this determination are $\Lambda_{QCD}$, and to a major extent the strange quark mass. From the positivity of $R_{su}$ there follows an upper bound on the latter: $\overline{m_{s}} (2 \; \mbox{GeV}) \;  \leq \; 121 \,(105) \; \mbox{MeV}$,  for $\Lambda_{QCD} = 330 \, (420) \; \mbox{MeV}\,.$ 


\newpage
\bigskip
\noindent
\section{Introduction}
\noindent
The ratio of strange to light-quark vacuum condensates, $R_{su} = \frac{<\bar{s} s>}{<\bar{q} q>}$, is a key QCD parameter measuring flavour SU(3) symmetry breaking in the vacuum \cite{PAGELS}. It is also an important quantity that enters in many QCD sum rule applications, e.g. baryon mass determinations, the Goldberger-Treiman discrepancy in $SU(3) \times SU(3)$, etc. \cite{REVIEW}-\cite{NS}. In addition, this ratio is related to two low energy constants of chiral perturbation theory \cite{CHPT}, which in turn determine the next-to-leading order corrections to the Gell-Mann, Oakes, Renner (GMOR) relation. As a result of this importance, many attempts have been made in the past to determine the numerical value of this ratio, as well as to improve its accuracy \cite{REVIEW}, \cite{CAD1}-\cite{JAMIN1}. Improvements in the QCD sector have been possible due to state of the art results for the relevant two-point functions at higher order in perturbation theory, as well as to a better understanding of how to deal with logarithmic quark-mass singularities. Better accuracy in the strange quark mass and in $\Lambda_{QCD}$ is still required. A serious limiting factor, though, has always been the lack of direct experimental information on the hadronic spectral functions entering the QCD sum rules used to extract $R_{su}$. While data on hadronic $\tau$-lepton decays has allowed for a simultaneous determination of the  (light) vector and axial-vector spectral functions, this is not yet possible for the scalar and pseudoscalar counterparts which determine $R_{su}$. Even if all scalar and pseudoscalar resonances were to be firmly established, a reconstruction of the hadronic spectral function would remain model-dependent to a large extent. In fact, inelasticity and non-resonant background are hard to model correctly. \\
In this paper we argue that it is valid to use QCD sum rules to determine the scalar and pseudoscalar two-point functions at zero momentum, which in turn determine the ratio $R_{su}$. These sum rules actually fix the  difference between the true $\psi_{(5)}(0)$ and its perturbative piece.
In an attempt to reduce systematic uncertainties from the hadronic sector
we introduce a new set of Finite Energy QCD sum rules (FESR) to estimate the scalar and pseudoscalar two-point functions at zero momentum . These FESR involve as integration kernel a second degree polynomial with two free parameters. These are determined by requiring the vanishing of the spectral function at the position of the first two resonances in each channel. As a result of this, the numerical importance of the hadronic contribution to the FESR is considerably reduced. In fact, it becomes roughly an order of magnitude smaller than the QCD counterpart. In addition, the latter turns out to be dominated by the purely perturbative QCD (PQCD) piece; the higher order in $m_s$ terms as well as the condensates add up to a negligible contribution as a result of partial cancellations. The results show a very good stability against changes in the upper limit of integration over a wide range of energies. Sensitivity to   $\Lambda_{QCD}$, and most particularly to $m_s$ remains somewhat high, and 
becomes the limiting factor in the accuracy that can be achieved for $R_{su}$. This is roughly at the 20 \% level. Nevertheless, with the hadronic uncertainties well under control in this approach, future  reduction of the errors in $m_s$ and $\Lambda_{QCD}$ will allow for a more accurate determination of $R_{su}$, almost free from hadronic systematic uncertainties.\\
The high sensitivity of the results for $\psi_{(5)}(0)$ and   
$R_{su}$ to the strange quark mass are used to derive an upper bound for this quantity from the requirement $R_{su} \geq 0$. Finally, the unphysical low energy constant of Chiral Perturbation Theory related to $R_{su}$ is estimated.

\section{Low energy theorem}
We introduce the correlator of vector and axial-vector divergences 

%Eq.1
\begin{equation}
\psi_{(5)} (q^{2})|_i^j   = i \, \int\; d^{4}  x \; e^{i q x} \; 
<|T(\partial^\mu J_{\mu}(x)|_i^j \;, \; \partial^\nu J_{\nu}^{\dagger}(0)|_i^j)|> \;,
\end{equation}

where $\partial^\mu J_{\mu}(x)|_i^j = (m_j \mp m_i) :\bar{q}_j(x) \,i \, (\gamma_{5})\, q_i(x):\;$ is the divergence of the vector (axial-vector) current. At zero momentum, a Ward identity relates the subtraction constants $\psi_{(5)}(0)|_i^j$ to the quark condensates \cite{PAGELS}, \cite{BROAD1}-\cite{BROAD2}, viz.

%Eq.2
\begin{equation}
\psi _{(5)}(0)|_i^j =-( m_j\mp m_i)
\left\langle [\overline{\psi }_j\psi _j\mp \overline{\psi }_i\psi
_i]\right\rangle \; .
\end{equation}

In the determination of $<\bar{s} s>$ we shall be using $i = u,d$ and $j = s$, as well as the approximations $m_s >> m_{u,d}$, $<\bar{u} u> \simeq <\bar{d} d>$. From the time-ordered product in Eq.(1) and using Wick's theorem one
would get normal-ordered operators in the low-energy theorem Eq.(2). However there are mass-singular quartic terms in  perturbative theory as well as
tadpole contributions. In fact, to lowest order in the $\overline{MS}$ scheme

%Eq.3
\begin{equation}
\psi_{(5)}^{\overline{MS}}(0)=-\frac{3}{4\pi^{2}}m_{s}^{4}(1+\ln\frac{\mu^{2}%
}{m_s^{2}})\ ,\label{4.2}%
\end{equation}

%Eq.4
\begin{equation}
\left\langle 0\left\vert m_{s}\bar{s}(0)s(0)\right\vert 0\right\rangle
^{\overline{MS}}= \frac{3}{4\pi^{2}}m_{s}^{4}(1+\ln\frac{\mu^{2}}{m_s^{2}
})\;.
\end{equation}

If the quark condensates in Eq. (2) would be
considered as minimally subtracted instead of normal-ordered, then the perturbative quartic mass corrections  would cancel \cite{BROAD2}-\cite{CAD3}, the low-energy theorem would make sense and the simple functional form of Eq.(2) would follow. All of this still holds after introducing gluonic corrections \cite{BROAD2}.\\

If the condensate is to be calculated from QCD sum rules, there is a subtle point concerning the renormalization of the operators in Eq.(1), first pointed out clearly in \cite{JAMIN1}, which we discuss in the following. In QCD the correlator $\psi_{(5)}(q^{2})$ is of the
general form

%Eq.5
\begin{equation}
\psi_{(5)}(q^{2})= A + B q^{2}+\tilde{\psi}_{(5)}(q^{2}) \;, \label{4.0}%
\end{equation}

where $A$ and $B$ are constants related to external renormalization, and $\tilde{\psi}_{(5)}(q^{2})$ is the two-point function without the first order polynomial, which has been factored out.  In the $\overline{MS}$ scheme of QCD perturbation theory, for instance, and at lowest
non-trivial order in the strong coupling constant, the  correlator at large $-q^2$ is given by

%Eq.6
\begin{eqnarray}
\psi_{(5)}^{\overline{MS}}(q^2) &\underset{-q^2 \rightarrow \infty}{=}&  m_{s}^{2} q^2 \frac{1}{16\pi^{2}} \left\{ 12- 6 L+ \frac{\alpha_s(q^2)}{\pi} \left[ \frac
{131}{2}-34 L+6 L^{2}-24 \zeta(3)\right]
 \right. \nonumber       \\[.3cm]
&+& \left. O(\alpha^2_s(q^2)) \phantom{\frac{1}{1}}\right\}
- m_{s}^{4} \;\frac{12}{16\pi^{2}}\; \left( 1-L\right) + O(\alpha_s m_s^4, m_s^6) \;, 
\end{eqnarray}

where $L=\ln\frac{-q^2}{\mu^{2}}$, and $\zeta(z)$ is the Riemann zeta-function. 
The correlator at zero momentum can be formally written in terms of a QCD Finite Energy Sum Rule (FESR), which follows from Cauchy's theorem in the complex energy (squared) plane (see Fig.1), i.e.

%Eq.7
\begin{eqnarray}
\psi_{(5)}(0)&=&
\int_{s_{th}}^{s_0}
\frac{ds}{s} \frac{1}{\pi} Im \;\psi_{(5)}(s)+ \frac{1}{2\pi i}
\oint_{C(|s_0|)}
\frac{ds}{s} \;\psi_{(5)}(s) \nonumber \\ [.3cm]
&\simeq&
\int_{s_{th}}^{s_0}
\frac{ds}{s}\frac{1}{\pi} Im \;\psi_{(5)}^{.}(s)+\frac{1}{2\pi i}
\oint_{C(|s_0|)}
\frac{ds}{s} \;\psi_{(5)}^{QCD}(s) \, ,\label{4.1a}
\end{eqnarray}

where $s_{th}$ is the hadronic threshold (e.g. $M_K^2$), and the contour integral is performed over a large circle where the exact $\psi_{(5)}(s)$ can  be safely replaced by its QCD counterpart $\psi_{(5)}^{QCD}(s)$.
To leading order in chiral-symmetry breaking, i.e. to order $\cal{O}$$(m_s^2)$, the constant $A$ in Eq.(5) vanishes, and the linear term in $q^2$ does not contribute to the integral in Eq.(7).

\begin{figure}
[ht]
\begin{center}
\includegraphics[height=1.9817in, width=1.8937in]
{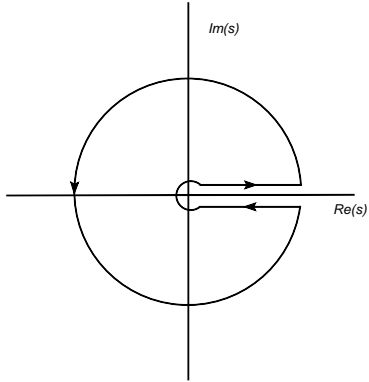}
\caption{Integration contour in the complex s-plane.}
\end{center}
\end{figure}

This means that in this case $\psi_{(5)}(0)$, as well as the non-normal ordered condensate in the low energy theorem, Eq.(2), can be determined unambiguously from a FESR. At the next order, i.e. keeping  terms of order $\cal{O}$ $(m_s^{4})$, and using Eq.(5) in Eq.(7) leads to

%Eq.8
\begin{equation}
\psi_{5}(0)=\int_{s_{th}}^{s_0}\; \frac{ds}{s}\frac{1}{\pi}\operatorname{Im}\psi_{5}(s)+\frac
{1}{2\pi i}{\displaystyle\oint\limits_{C(|s_0|)}}
\frac{ds}{s} \; \tilde{\psi}_{5}^{\overline{MS}}(s) 
\end{equation}

where $\tilde{\psi}_{5}^{QCD}(s)$ is defined in Eq.(5) . The sum rule then relates the non-normal ordered condensate to an integral over the hadronic spectral function and a contour integral over the non-polynomial part of the QCD correlator $\tilde{\psi}_{5}^{QCD}(s)$.
Whereas the product of quark mass times normal-ordered condensate is a renormalization
invariant quantity, this is, however, no longer true for the
non-normal-ordered condensate. Only at leading order in $m_{s}^{2}$ is the quark condensate directly related to a physical quantity. In full QCD, however, its value, just like the QCD coupling and the quark masses, depends on the renormalization scale and on the renormalization scheme employed.

Numerically, the quartic mass terms are potentially relevant only for the determination of the strange quark mass. In fact, we find  a-posteriori 
that the corrections of order $m_{s}^{4}$ to $\psi_{(5)}(0)$ are at the level of
only $(1-2)\%$ of the leading terms. Hence,  the subtleties of
renormalization discussed above are largely academic.

\section{The ratio $R_{su}$}
One possible way of determining the ratio $R_{su}$ is to use the auxiliary ratio

%Eq.9
\begin{equation}
R_{AA} \equiv \frac {\psi_5(0)|^s_u} {\psi_5(0)|^d_u}
= \frac{1}{2} \;
\frac{m_s+m_u}{m_u+m_d} \;(1 + \frac{<\bar{s}s>}{<\bar{u}u>}) \; ,
\end{equation}

where $<\bar{u}u>\, \simeq <\bar{d}d>$ will be assumed in the sequel. In fact, if the subtraction constants and the quark masses are determined independently, e.g. from QCD sum rules, then $R_{su}$ follows. Using current values of the quark masses \cite{PDG} gives

%eq.10
\begin{equation}
R_{su}\equiv \frac{<\bar{s}s>}{<\bar{u}u>} \simeq 0.15\; R_{AA} - 1 \; .
\end{equation}

Since $R_{AA}$ is expected from current algebra to be of order $\cal{O}$(10), this method would result in a very large uncertainty in $R_{su}$ unless the subtraction constants were to be determined with extreme accuracy. Due to this, an alternative procedure, first proposed in \cite{CAD2}, consists in using instead the ratio

%eq.11
\begin{equation}
R_{VA}\equiv \frac{\psi( 0) _u^s}{\psi _5(0) _u^s}\ ,
\end{equation}
which leads to

%Eq.12
\begin{equation}
R_{su} \equiv
\frac{\left\langle \overline{s}s\right\rangle }{\left\langle \overline{u}
u\right\rangle }\simeq \frac{1+R_{VA}}{1-R_{VA}} \; .
\end{equation}

This method was used in \cite{CAD1} to obtain both subtraction constants  from Laplace transform QCD sum rules to four loops with the results

%Eq.13
\begin{equation}
\psi_5(0)|^s_u =  (3.35 \pm 0.25) \times 10^{-3} \; \mbox{GeV}^4 \;,
\end{equation}

%Eq.14
\begin{equation}
\psi(0)|^s_u= - (1.06 \pm 0.21) \times 10^{-3} \; \mbox{GeV}^4 \;,
\end{equation}

%Eq.15
\begin{equation}
R_{su}\equiv \frac{<\bar{s}s>}{<\bar{u}u>} =  0.5 \pm 0.1      \;,
\end{equation}

and the following value of the invariant strange-quark mass

%Eq.16
\begin{equation}
 \widehat{m}_s = 140 \pm 10 \;\mbox{MeV} \; ,
\end{equation}

for $\Lambda_{QCD}$ in the range $\Lambda_{QCD}\simeq 300-350\;\mbox{MeV}$. Not included in the above errors are the uncertainties due to  hadronic spectral function modelling, which could be large \cite{PAVER1}.\\

The connection between the quark condensate ratio $R_{su}$ determined from QCD sum rules, e.g. through
Eq.(12), and the one entering chiral perturbation theory has been discussed in \cite{JAMIN1}. In the framework of the latter, $R_{su}$ depends on an unphysical low-energy constant $H^r_2$ through the relation

%eq.17
\begin{equation}
R_{su}\equiv \frac{<\bar{s} s>}{<\bar{q} q>} = 1 + 3 \mu_\pi - 2 \mu_K - \mu_\eta + \frac{8}{f_\pi^2} (M_K^2 - M_\pi^2) (2 L^r_8 + H^r_2) \; ,
\end{equation}

where $<\bar{q} q>$ is the average of the up- and down quark condensates, $L^r_8$ is a (physical) low-energy constant in the  chiral Lagrangian to next-to-leading order \cite{BERN}, and 

%eq.18 
\begin{equation}
\mu_P = \frac{M_P^2}{32 \pi^2 f_\pi^2} \ln \frac{M_P^2}{\nu_\chi^2} \;
\end{equation}

with $\nu_\chi$ the chiral renormalization scale. The constant $L^r_8$ at a scale equal to the rho-meson mass has been estimated in chiral perturbation theory to next-to-leading order  with the result \cite{JAMIN1}

%eq.19
\begin{equation}
L^r_8(\nu_\chi = M_\rho) = (0.88 \pm 0.24) \times 10^{-3} \;,
\end{equation}

while a determination at order $\cal{O}$$(p^6)$ gives \cite{P6} $L^r_8(\nu_\chi = M_\rho) = (0.62 \pm 0.20) \times 10^{-3}$. The unphysical low energy constant $H^r_2$ has been estimated in \cite{JAMIN1} as

%eq.20
\begin{equation}
H^r_2 (\nu_\chi = M_\rho) = - (3.4 \pm 1.5) \times 10^{-3} \;.
\end{equation}

Both low energy constants determine the size of the next-to-leading order chiral corrections to the GMOR relations in $SU(2)\times SU(2)$ and $SU(3)\times SU(3)$,  $\delta_\pi$ and $\delta_K$ respectively, defined as

%eq.21
\begin{equation}
(m_u + m_d) <\bar{u}u + \bar{d}d> = - 2 f_\pi^2 M_\pi^2 (1 - \delta_\pi) \;,
\end{equation}

%eq.22
\begin{equation}
m_s <\bar{s}s> (1 + \frac{1}{R_{su}}) = - 2 f_K^2 M_K^2 (1 - \delta_K) \;,
\end{equation}

where the physical values of the pseudoscalar decay constants are $f_\pi = 92.4 \pm 0.26 \;\mbox{MeV}$, and $f_K/f_\pi = 1.22 \pm 0.01$ \cite{PDG}. To next-to-leading order one has \cite{BERN}

%eq.23
\begin{equation}
\delta_\pi = 4 \frac{M_\pi^2}{f_\pi^2} ( 2 L^r_8 - H^r_2) \;\;\;\; and\;\;\;\; \delta_K = \frac{M_K^2}{M_\pi^2} \,\delta_\pi \;.
\end{equation}

\section{Finite Energy QCD Sum Rules}

We consider first the pseudoscalar correlator, $\psi_5(q^2)$, which exhibits a pole and a cut in the complex energy (squared) plane. Cuachy's theorem reads

%eq.24
\begin{equation}
\psi_5(0) = 2 f_K^2 M_K^2 + \frac{1}{2 \pi i} \oint_{C} \frac{ds}{s}\;
\psi_5(s) \;,
\end{equation}

where the closed contour $C$ comprises the cut across the real axis and the circle of radius $|s_0|$ (see Fig. 1).
Introducing an integration kernel of the form

%eq.25
\begin{equation}
\Delta_5(s) = 1 - a_{05} \;s - a_{15}\; s^2 \;,
\end{equation}

where $a_{05}$, and $a_{15}$ are free parameters, the two-point function at zero momentum becomes

%eq.26
\begin{eqnarray}
\psi_5 (0) &=& 2 f_K^2 M_K^2 \; \Delta_5 (M_K^2) + \frac{1}{\pi} \int_{s_{th}}^{s_0} \frac{ds}{s} \;\Delta_5 (s)\; Im \;\psi_5 (s)|_{RES} \nonumber \\ [.3cm]
&+&
\frac{1}{2 \pi i} \oint_{C(|s_0|)} \;\frac{ds}{s}\; \Delta_5(s) \; \psi_{(5)}(s)|_{QCD} \;.
\end{eqnarray}

The free parameters $a_{05}$, and $a_{15}$ will be chosen in such a way that $\Delta_5(M_1^2) = \Delta_5(M_2^2) = 0$, where
$M_{1,2}$ are the masses of the two radial excitations of the kaon. This procedure will reduce considerably the numerical importance of the resonance contribution to $\psi_5(0)$, thus reducing the systematic uncertainties that plague the hadronic sector. For the scalar two-point function at zero momentum one finds

%eq.27
\begin{equation}
\psi(0) = \frac{1}{\pi} \int_{s_{th}}^{s_{0}} \frac{ds}{s} \;\Delta(s)\; Im \;\psi(s)|_{RES} +
\frac{1}{2 \pi i} \oint_{C(|s_0|)} \frac{ds}{s}\; \Delta(s)\; \psi(s)|_{QCD} \;,
\end{equation}

where $\Delta(s)$ is a second degree polynomial as in Eq. (25), and it will also be constrained to vanish at the position of the two resonances in the scalar channel.

The two-point function $\psi_{(5)}(q^2)$ has been known  in PQCD up to four-loops for quite some time \cite{PQCD4}. Recently, the PQCD second derivative of $\psi_{(5)}(q^2)$ to five loops  has been computed in \cite{PQCD5}. Integrating this result twice gives the five-loop expression for the two-point function up to polynomial terms . The latter do not contribute to the integrals around the circle in the s-plane. The remaining terms in the QCD expression for $\psi_{(5)}(q^2)$, i.e. the higher orders in $\overline{m}_s$ and the quark and vacuum condensate contributions, may be found in \cite{CAD3}. To compute the QCD contribution we define

%eq.28
\begin{equation}
\delta_{(5)}(s_0)|_{QCD} \equiv \frac{1}{2 \pi i} \oint_{C(|s_0|)} \frac{ds}{s}\; \Delta_{(5)}(s)\; \psi_{(5)}(s)|_{QCD} \;,
\end{equation}

where $\psi_5(s)|_{QCD} = \psi(s)|_{QCD}$ for the purely gluonic piece, but $\delta_5(s_0) \neq \delta (s_0)$ on account of $\Delta_5(s) \neq \Delta(s)$
($a_{05}\neq a_0$ and $a_{15}\neq a_1$). Using the expression for $\psi_{(5)}(q^2)$ to five-loop order in Eq. (28) and performing the integration around the circle in the complex s-plane we find the following purely gluonic results

%eq.29
\begin{equation}
\delta_{(5)}|_{1LOOP} = - \frac{1}{16 \pi^2} \;\overline{m}_s^2(s_0)\; \left[6 \,s_0 - 3\, a_{0(5)} \, s_0^2 - 2\, a_{1(5)}\, s_0^3 \right] \:,
\end{equation}

%eq.30
\begin{equation}
\delta_{(5)}|_{2LOOP} = - \frac{1}{16 \pi^2} \;\overline{m}_s^2(s_0)\; \frac{\alpha_s(s_0)}{\pi}\; \left[46 \, s_0 - 20\, a_{0(5)}\, s_0^2 - \frac{38}{3}\; a_{1(5)}\; s_0^3 \right] \:,
\end{equation}

%eq.31
\begin{eqnarray}
\delta_{(5)}|_{3LOOP} &=& - \frac{1}{16 \pi^2}\; \overline{m}_s^2(s_0)\; [\frac{\alpha_s(s_0)}{\pi}]^2 \; \left\{ s_0 \left[ \frac{9631}{24} - 105\; \zeta(3)  \right. 
\right.  \nonumber \\ [.3cm]
&+& \left. \left. 190  -51 \;(\frac{\pi^2}{6} - 1) \right]
 - a_{0(5)} \frac{s_0^2}{2} \left[\frac{9631}{24} - 105 \; \zeta(3) + 95 
 \right. \right. \nonumber \\ [.3cm]
  &-& \left. \left. 51 \; (\frac{\pi^2}{6} - \frac{1}{4}) \right] 
 -  a_{1(5)} \frac{s_0^3}{3} \left[\frac{9631}{24} - 105 \; \zeta(3) + \frac{190}{3} \right. \right. \nonumber \\ [.3cm]
 &-& \left. \left. 51 \; (\frac{\pi^2}{6} - \frac{1}{9}) \right]  
    \right\} \:,
\end{eqnarray}

%eq.32
\begin{eqnarray}
\delta_{(5)}|_{4LOOP} &=& - \frac{1}{16 \pi^2}\; \overline{m}_s^2(s_0)\; [\frac{\alpha_s(s_0)}{\pi}]^3 \; \left\{ s_0 \left[ A_1 + 12 ( \frac{4781}{18} - \frac{475}{8}\; \zeta(3))   \right. 
\right.  \nonumber \\ [.3cm]
&-& \left. \left. (1374 + \frac{663}{2})   \;(\frac{\pi^2}{6} - 1) \right]
 - a_{0(5)} \frac{s_0^2}{2} \left[A_1 + 6 (\frac{4781}{18} - \frac{475}{8} 
 \right. \right. \nonumber \\ [.3cm]
  &\times& \left. \left. \zeta(3)) - (1374 + \frac{663}{4}) \; (\frac{\pi^2}{6} - \frac{1}{4}) \right] 
 -  a_{1(5)} \frac{s_0^3}{3} \left[A_1 + 4 (\frac{4781}{18} 
\right. \right. \nonumber \\ [.3cm] 
   &-& \left. \left. \frac{475}{8} \; \zeta(3)) - (1374 + \frac{221}{2}) \; (\frac{\pi^2}{6} - \frac{1}{9}) \right]  
    \right\} \:,
\end{eqnarray}

%eq.33
\begin{eqnarray}
\delta_{(5)}|_{5LOOP} &=& - \frac{1}{16 \pi^2}\; \overline{m}_s^2(s_0)\; [\frac{\alpha_s(s_0)}{\pi}]^4 \; \left\{ s_0 \phantom{\frac{1}{1}}\left[ H_1 - 2 H_2
- (6 H_3 - 24 H_4) \phantom{\frac{1}{1}} \right.\right.  \nonumber \\ [.3cm]
&\times& \left. \left. (\frac{\pi^2}{6} - 1) + 120 \,H_5\; (\frac{\pi^4}{120} - \frac{\pi^2}{6} + 1) \right]
- a_{0(5)} \frac{s_0^2}{2} \left[H_1 -  H_2 \phantom{\frac{1}{1}}  \right.\right.  \nonumber \\ [.3cm]
&-& \left. \left. (6 H_3 - 12 H_4)\; (\frac{\pi^2}{6} - \frac{1}{4})+120 \, H_5\; (\frac{\pi^4}{120} - \frac{\pi^2}{24} + \frac{1}{16}) \right]
 \right.   \nonumber \\ [.3cm]
&-& \left.    a_{1(5)} \frac{s_0^3}{3} \left[H_1 - \frac{2}{3} H_2-(6 H_3 - 8 H_4)\; (\frac{\pi^2}{6} - \frac{1}{9})+ 120 \, H_5      \right. \right.  \nonumber \\ [.3cm]
&\times& \left. \left.  \;(\frac{\pi^4}{120} -\frac{\pi^2}{54} + \frac{1}{81}) \right]\right\} \;
\end{eqnarray}

where for three quark flavours $A_1 = 2795.0778$, $H_1 = 33 532.30$, $H_2 = - 15 230.645$, $H_3 = 3 962.4549$, $H_4 = - 534.05208$, and $H_5 = 24.171875$. The constants $H_i$ enter in the expression of the two-point functions to five loops as

%eq.34
\begin{equation}
\psi_{(5)}(q^2)|_{5LOOP} = \frac{1}{16 \pi^2} \overline{m}_s^2 (- q^2)\;
[\frac{\alpha_s(s_0)}{\pi}]^4 \;\sum_{i=1}^{5} H_i \, L^i \;,
\end{equation}

where $L = \ln (-q^2/\mu^2)$, and the above expansion is up to (unknown) terms not multiplying logarithms, which do not contribute to $\delta_{(5)}$. The remaining QCD contributions to $\delta_{(5)}$ (higher order in $m_s$ and vacuum condensates) have also been calculated, but their total contribution is at the level of $1-2 \%$ of the sum of the gluonic terms in the wide range $s_0 \simeq 2 - 6 \;\mbox{GeV}^2$.\\

Turning to the hadronic sector, the spectral function in the pseudoscalar channel, $Im \,\psi_5(s)|_{HAD}$ involves the kaon pole, plus two radial excitations, the K(1460) and K(1830) both with widths of about 250 MeV. We follow the procedure outlined in \cite{CAD4}, where the resonance part of the spectral function is written as a linear combination of two Breit-Wigner forms normalized at threshold according to chiral perturbation theory. The latter incorporates the resonant sub-channel $K^*(892)-\pi$ which is important due to the narrow width of the $K^*(892)$. This gives

%eq.35
\begin{eqnarray}
\delta_{5}(s_0)|_{HAD} &=&  2 f_K^2 M_K^2 \; \Delta_5 (M_K^2) + \frac{1}{\pi} \int_0^{s_0} \frac{ds}{s} \;\Delta_5 (s)\; Im \;\psi_5 (s)|_{RES} \nonumber \\ [.3cm]
&\equiv&  \delta_{5}(s_0)|_{POLE} + \delta_{5}(s_0)|_{RES} \;.
\end{eqnarray}

For the scalar channel there is experimental data on $K \pi$ phase shifts \cite{KPI} that can be used to reconstruct the spectral function

%eq.36
\begin{equation}
\frac {1}{\pi} Im\; \psi \left( s\right) =\frac {3}{32\pi ^2}
\frac{\sqrt{\left( s- s_{+}\right) \left( s- s_{-}
\right)}}{s} \; \left| d(s)\right|^2 \; ,
\end{equation}

where $s_{\pm} = (M_K \pm M_\pi)^2$, and $d(s)$ is the scalar form
factor. One can use the method of \cite{PAVER1}, based on the Omn\`{e}s
representation, to relate $d(s)$ to the experimental phase shifts.  A posteriori, the numerical importance of the resonance contribution to 
$\delta_{(5)}(s_0)|_{HAD}$ is one order of magnitude smaller than the gluonic contributions on account of the integration kernel $\Delta_{(5)}(s)$. Hence, a simpler parametrization in terms of two Breit-Wigner forms, properly normalized at threshold with $ |d(s_+)| \simeq 0.3 \;\mbox{GeV}^2$, is equally acceptable. We thus include the $K^*_0(1430)$ and the $K^*_0(1950)$ with masses and widths $M_1 = 1.4\; \mbox{GeV}$, $\Gamma_1 = 290 \pm 21\; \mbox{MeV}$, and $M_2 = 1.94\; \mbox{GeV}$, $\Gamma_2 = 201 \pm 86\; \mbox{MeV}$.\\
The function $\delta(s_0)$ in this channel can then be written as

%eq.37
\begin{equation}
\delta(s_0)|_{HAD} =   \frac{1}{\pi} \int_0^{s_0} \frac{ds}{s} \;\Delta(s)\; Im \;\psi (s)|_{RES} \equiv \delta(s_0)|_{RES} \;.
\end{equation}

Requiring  $\Delta_{(5)}(s)$ to vanish at resonance determines $a_{0(5)}$ and $a_{1(5)}$ with the result

%eq.38
\begin{eqnarray}
\begin{array}{lcl}
a_0 = 0.777  \; \mbox{GeV}^{-2} \; \; \;\;a_1 = - 0.136 \; \mbox{GeV}^{-4}\\[.3cm]
a_{05} = 0.768  \; \mbox{GeV}^{-2} \; \;  a_{15} = - 0.140 \;\; \mbox{GeV}^{-4} \;.
\end{array}
\end{eqnarray}

The values of these coefficients are very similar on account of the similarity between the scalar and pseudoscalar resonance masses.
\section{Results}

In order to compute the QCD contribution to the scalar and pseudoscalar two-point functions at zero momentum we need as input the invariant strange-quark mass $\hat{m_s}$ and the QCD scale $\Lambda_{QCD}$, which are strongly correlated. To obtain the running quark mass and strong coupling constant it is only necessary to use the four-loop expressions, which for three quark flavours are

%eq.39
\begin{eqnarray}
\overline{m}_s\left( Q^2\right) &=&\frac{\widehat{m}_s}{\left( \frac
12L\right) ^{\frac 49}}\left\{ 1+\left( 290-256LL\right) \frac 1{729}\frac
1L\right.  \nonumber \\
&&\nonumber \\
&&+ \left[  \frac{550435}{1062882}-\frac{80}{729}\;\zeta \left(
3\right) \right. \nonumber \\
&&\nonumber \\
&&-\left( 388736LL-106496LL^2\right)\left.  \frac 1{531441} \right] \frac
1{L^2}  \nonumber \\
&&\nonumber \\
&&+\left[  -\frac{126940037}{1162261467}-\frac{256}{177147}\;\beta _4+
\frac{128}{19683}\;\gamma _4+\frac{7520}{531441}\;\zeta \left( 3\right) 
\right. \nonumber \\
&&\nonumber \\
&& + \left( -\frac{611418176}{387420489}+
\frac{112640}{531441}\;\zeta \left(
3\right) \right) LL+\frac{335011840}{387420489}LL^2  \nonumber \\
&&\nonumber \\
&&- \left. \frac{149946368}{1162261467}LL^3  \right] \frac 1{L^3}
\left. +
{\cal O}\left( \frac {1}{L^4}\right) \right\} \ ,  
\end{eqnarray}

%eq.40
\begin{eqnarray}
\frac{\alpha_{s}(s_{0})}{\pi} &=&
\frac{\alpha^{(1)}_{s}(s_{0})}{\pi}
+ \Biggl (\frac{\alpha^{(1)}_{s}(s_{0})}{\pi}\Biggr )^{2}
\Biggl (\frac{- \beta_{2}}{\beta_{1}} {\rm ln} L \Biggr ) \nonumber \\ [.3cm]
&+& \Biggl (\frac{\alpha^{(1)}_{s}(s_{0})}{\pi}\Biggr )^{3} 
\Biggl (\frac{\beta_{2}^{2}}{\beta_{1}^{2}} ( {\rm ln}^{2} L -
{\rm ln} L -1) + \frac{ \beta_{3}}{\beta_{1}} \Biggr ) \nonumber \\ [.3cm]
&+& {\cal{O}} (1/L^4) \;,
\end{eqnarray}

where
%Eq.41
\begin{equation}
\frac{\alpha^{(1)}_{s}(s_{0})}{\pi} \equiv\frac{- 2}{\beta_{1} L}\; ,
\end{equation}

$L = \ln (s_0/\Lambda_{QCD}^2)$, $LL = \ln L$, $\beta_1 = - 9/2$, $\beta_2 = - 8$, $\beta_3 = - 3863/192$, and

%Eq.42
\begin{equation}
\beta _4=-\frac{281198}{4608}-\frac{890}{32}\;\zeta \left( 3\right) \ , 
\end{equation}

with $\gamma_4 = 88.5258$ \cite{G4}, and $\widehat{m}_j$ is the invariant
quark-mass.
The terms of order ${\cal O}$$\left( \frac {1}{L^4}\right) $ above
are known up to a constant not determined by the renormalization group.
However, we have checked that our final results
are essentially insensitive to the inclusion of terms of this order in $\alpha_s$ and $\overline{m_s}$.\\

Since we are dealing with three quark flavours, it is simpler to determine $\Lambda_{QCD}$ from the strong coupling obtained from $\tau$-decay \cite{PDG}, \cite{ALEPH}: $\alpha_s(M_\tau^2) = 0.31 - 0.36$, which gives $\Lambda_{QCD} = 330 - 420\; \mbox{MeV}$. Recent determinations of the strange quark mass from various QCD sum rules \cite{CAD1}, \cite{PQCD5}, \cite{ms} give values in the range $\overline{m_s} (2 \; \mbox{GeV}) \simeq 80 - 130 \; \mbox{MeV}$, which translates into $\hat{m_s} \simeq 100 - 170 \;\mbox{MeV}$ after using the above values of $\Lambda_{QCD}$. The two-point functions at zero momentum are given by

%eq.43
\begin{equation}
\psi(0)|^s_u = \delta|_{RES}(s_0) + \delta|_{QCD}(s_0) \;,
\end{equation}

%eq.44
\begin{equation}
\psi_5(0)|^s_u = \delta_5|_{POLE}(s_0) + \delta_5|_{RES}(s_0) + \delta_5|_{QCD}(s_0) \; ,
\end{equation}

where the various $\delta's$ above are computed from Eqs.(29)-(33),(35) and (37); as mentioned earlier, the QCD contributions to $\delta_{(5)}(s_0)$ from higher orders in $m_s$ and from the vacuum condensates add up (due to partial cancellations) to less than 1\% of the sum of the gluonic terms. In Fig.1 we show the results for $\psi_5(s_0)$ (curve (a)), and $\psi(0)$ (curve b) as a function of $s_0$ for the reference value of the invariant quark mass $\hat{m_s} = 100 \; \mbox{MeV}$ ($\overline{m_s}(2 \;\mbox{GeV}) \simeq 80 \;\mbox{MeV}$), and $\Lambda_{QCD} = 330 \; \mbox{MeV}$. As seen from this figure the results are fairly stable in the wide region $s_0 \simeq 2 - 6 \;\mbox{GeV}^2$; a similar stability is obtained for $\Lambda_{QCD} = 420 \; \mbox{MeV}$. For values in the range $\Lambda_{QCD} = 330 - 420\; \mbox{MeV}$ we find

%eq.45
\begin{equation}
\psi_5(0) = (0.39 \, \pm\, 0.03) \times \, 10^{-2} \; \mbox{GeV}^4 \;\;\; (\hat{m_s} = 100 \; \mbox{MeV}) \;,
\end{equation}

%eq.46
\begin{equation}
\psi(0) = - (0.95 \, \pm\, 0.25) \times \, 10^{-3} \; \mbox{GeV}^4 \;\;\; (\hat{m_s} = 100 \; \mbox{MeV}) \;,
\end{equation}

which using Eqs.(11)-(12) leads to

%eq.47
\begin{equation}
R_{su} \equiv
\frac{\left\langle \overline{s}s\right\rangle }{\left\langle \overline{u}
u\right\rangle } = 0.6 \, \pm 0.1 \;\;\; (\hat{m_s} = 100 \; \mbox{MeV}) \; .
\end{equation}

\begin{figure}[ht]
\begin{center}
\includegraphics[width=\columnwidth]{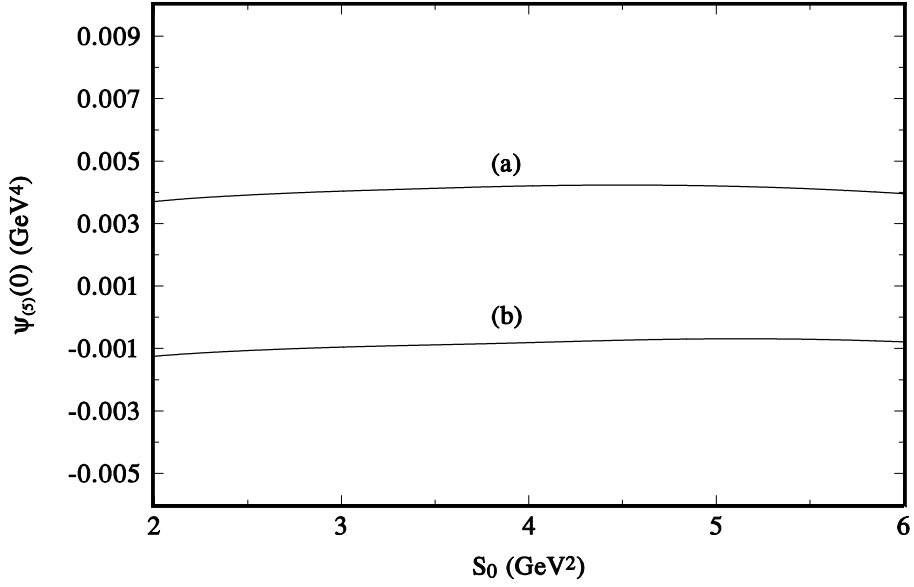}
\caption{The two-point functions at zero momentum, $\psi_5(0)$, curve (a), and $\psi(0)$, curve (b), as a function of $s_0$, for $\hat{m_s} = 100 \; \mbox{MeV}$, and $\Lambda_{QCD} = 330 \; \mbox{MeV}$.}
\end{center}
\end{figure}

The uncertainties above are entirely due to the uncertainty in $\Lambda_{QCD}$, as $\hat{m_s}$ has been kept fixed at the indicated reference value. The ratio $R_{su}$ exhibits a stronger sensitivity to the value of the strange quark mass, as this enters in the PQCD expression of $\psi_{(5)}$ as an overall multiplicative factor $\hat{m_s}^2$. \\
\newpage
Given the similarity between the integration kernels in the scalar and pseudoscalar channels, $\Delta(s) \;\simeq\; \Delta_5(s)$, or $\delta|_{QCD}(s_0) \;\simeq\; \delta_5|_{QCD}(s_0)$, it is possible to obtain an approximate expression for $R_{su}$ as a function of $\hat{m_s}$ as follows. From Eqs.(12) and (43)-(44) one has

%eq.48
\begin{eqnarray}
R_{su} &=& \frac{\delta_5|_{POLE} \;+\; \delta_5|_{RES} \;+\; \delta_5|_{QCD} \;+\; \delta|_{RES} \;+\; \delta|_{QCD}}
{\delta_5|_{POLE} \;+\; \delta_5|_{RES} \; - \; \delta_{RES} \;+\delta_5|_{QCD} \;-\; \delta|_{QCD}} \nonumber \\ [.3cm]
&\simeq& \Bigg[ \frac{\delta_5|_{POLE} \;+\; \delta_5|_{RES} \; +
\delta|_{RES}}{\delta_5|_{POLE} \;+\; \delta_5|_{RES} \; -
\delta|_{RES}}\Bigg] \;+\; \Bigg[\frac{\delta_5|_{QCD}\;+\;\delta|_{QCD}}
{\delta_5|_{POLE} \;+\; \delta_5|_{RES} \; -\;
\delta|_{RES}}\Bigg]\nonumber \\ [.3cm]
&\equiv& A \; + B(\Lambda_{QCD})\; \Big[ \frac{ \hat{m_s} (\mbox{MeV})}{100\; \mbox{MeV}}\Big]^2 \;,
\end{eqnarray}

where we approximated $[\delta_5|_{QCD} - \delta|_{QCD}] \simeq 0$ in the denominator of the above ratio, $A \simeq 1.15$ is basically constant in the wide range $s_0 = 2 \,-\,6\;\mbox{GeV}^2$, and $B(\Lambda_{QCD}) = - 0.44 \,( -0.68)$ for $\Lambda_{QCD} = 330 \,(420)\; \mbox{MeV}$, respectively. This formula is accurate to within 2 - 3 \%, and it allows for a quick estimate of $R_{su}$ for other values of the invariant strange quark mass. It also gives an upper bound for this mass from the fact that $R_{su} \geq 0$, viz.

%eq.49
\begin{eqnarray}
\hat{m_{s}} \;  \leq \;\Bigg\{ 
\begin{array}{lcl}
162  \; \mbox{MeV} \; \; \; (\Lambda_{QCD} = 330\; \mbox{MeV})\\[.3cm]
130  \; \mbox{MeV} \; \; \; (\Lambda_{QCD} = 420 \; \mbox{MeV}) \;.
\end{array}
\end{eqnarray}

These bounds translate into the following bounds for the running strange quark mass at a scale of 2 GeV

%eq.50
\begin{eqnarray}
\overline{m_{s}} (2 \; \mbox{GeV}) \;  \leq \;\Bigg\{ 
\begin{array}{lcl}
121  \; \mbox{MeV} \; \; \; (\Lambda_{QCD} = 330 \; \mbox{MeV})\\[.3cm]
105  \; \mbox{MeV} \; \; \; (\Lambda_{QCD} = 420 \; \mbox{MeV}) \;.\end{array}
\end{eqnarray}
\begin{figure}[ht]
\begin{center}
\includegraphics[width=\columnwidth]{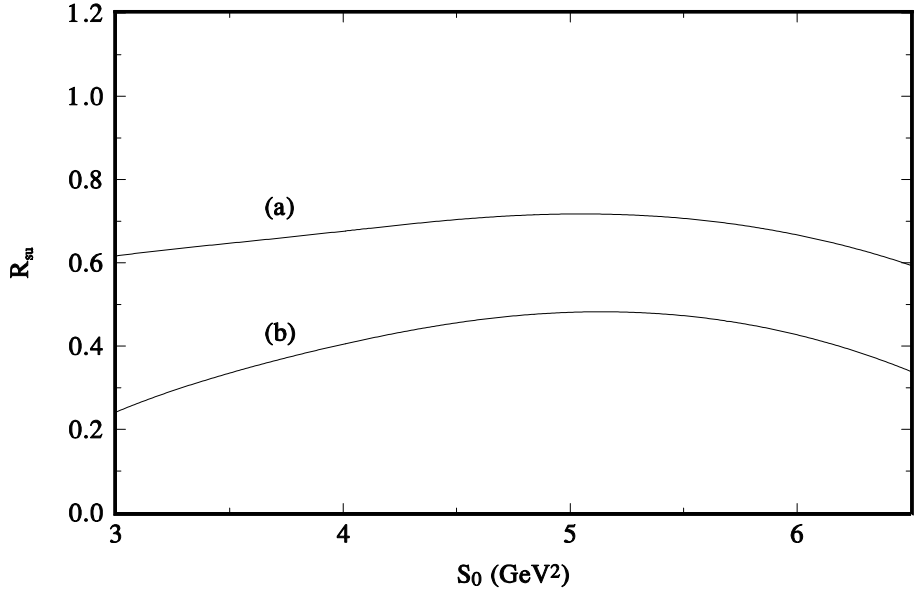}
\caption{The ratio $R_{su}$ as a function of $s_0$, for $\hat{m_s} = 100 \; \mbox{MeV}$,  $\Lambda_{QCD} = 330 \; \mbox{MeV}$, curve (a), and $\Lambda_{QCD} = 420 \; \mbox{MeV}$, curve (b).}
\end{center}
\end{figure}
These results are in line with recent determinations from QCD sum rules \cite{CAD1}, \cite{PQCD5}, \cite{ms}, as well as Lattice QCD \cite{LATT}. However, in making comparisons, the strong correlation between $\hat{m_s}$ and $\Lambda_{QCD}$ should be kept in mind. In particular, older determinations giving higher values of $m_s$ used mostly $\Lambda_{QCD} \simeq 100 - 250 \;\mbox{MeV}$.\\

We now turn to the implications of these results for chiral perturbation theory, as outlined in Section 2. Inserting our result for $R_{su}$, Eq.(47), in Eq. (17), and using Eqs. (18) and (19) gives the following prediction for the low energy constant $H^r_2$

%eq.51
\begin{equation}
H^r_2 = - (4.3 \pm 1.3) \times 10^{-3} \; ,
\end{equation}

where the range $f_\pi = 82 \; -\; 92 \;\mbox{MeV}$ was used, to take into account uncertainties from higher orders in the chiral expansion \cite{JAMIN1}.
\newpage
Using this result together with Eq. (19) in Eq. (23), the next-to-leading order corrections to the GMOR relation  become

%eq.52
\begin{eqnarray}
\begin{array}{lcl}
\delta_\pi = 0.04 \pm 0.02\\[.3cm]
\delta_K = 0.5 \pm 0.2\;,
\end{array}
\end{eqnarray}

in good agreement with \cite{JAMIN1}.

\section{Conclusions}

We have argued that it is legitimate to use QCD sum rules to determine the scalar and pseudoscalar two-point functions at zero momentum. These sum rules actually fix the  difference between the true $\psi_{(5)}(0)$ and its perturbative piece.
Approaches based on traditional QCD sum rules, e.g. Laplace transform sum rules, are affected by  uncontrollable systematic uncertainties in the reconstruction of hadronic resonance spectral functions. To minimize these uncertainties we have introduced new Finite Energy QCD sum rules (FESR) involving an integration kernel in the form of a second degree polynomial with two free parameters. Requiring  the hadronic spectral function to vanish at the position of the first two resonances determines these constants, and reduces the importance of this contribution to the FESR by one order of magnitude. This makes $\psi_{(5)}(0)$ dependent mostly on the strange quark mass, and to a lesser extent on $\Lambda_{QCD}$. The dependence on the radius $s_0$ of the integration contour in the complex energy plane is very mild, with the results for $\psi_{(5)}(0)$ showing very good stability in the wide range $s_0 \simeq 2 - 6 \;\mbox{GeV}^2$. Our results for the scalar and pseudoscalar correlators at zero momentum, Eqs.(45)-(46), as well as for the ratio $R_{su}$, Eq. (47), are in broad agreement with most previous determinations based on traditional QCD sum rules (Laplace, FESR) \cite{REVIEW}, \cite{CAD1}-\cite{JAMIN1}. However, it should be kept in mind that many of the old dterminations used much lower values of $\Lambda_{QCD}$, and somewhat higher values of $\hat{m_s}$. In addition, they used available PQCD results at the time, which were limited to two-, three- or at most four-loop order, many were affected by logarithmic quark-mass singularities, and  by uncontrollable systematic hadronic uncertainties in all cases. Our result for $R_{su}$ is somewhat smaller than one from a recent lattice determination \cite{PENN}. The bound obtained for $m_s$, Eqs.(49)-(50), is in good agreement with recent results from QCD sum rules \cite{ms}, as well as lattice QCD \cite{LATT}, which point to values smaller than in the past. Future improvement in the accuracy of $\Lambda_{QCD}$, and particularly in that of $m_s$, will allow for a more precise determination of the scalar and pseudoscalar correlators at zero momentum, and hence of $R_{su}$, almost free of systematic hadronic uncertainties. This becomes possible due to the introduction of the integration kernel, Eq.(25), in the FESR. It should be mentioned that this kernel, while vanishing at the resonance peaks, it does not vanish at the point $s = s_0$ where the integration circle in the complex energy plane meets the real axis. It is known that in some applications of FESR, e.g. in tau-decay, perturbative QCD does not seem to hold close to the real axis; this has led to the proposal of weighted FESR with weight functions vanishing at $s = s_0$ \cite{MALT}. In the application discussed here, though, this problem does not seem to arise. In fact, the perturbative expansion appears to converge very well, and the stability region is unusually broad, extending well above standard values. The introduction of an additional integration kernel vanishing at $s = s_0$ would not seem to provide any additional advantages. In any case, we have confirmed this by an explicit calculation. The subtraction constants $\psi_{(5)}(0)$ remain essentially the same if a weighted kernel is added.

\end{document}